\def\stackunder#1#2{\mathrel{\mathop{#2}\limits_{#1}}}
\def\dfrac#1#2{{\displaystyle {#1 \over #2}}}
\documentstyle[aps,preprint]{revtex}
\begin{document}
\title{The gravitational and electrostatic fields far from an isolated 
Einstein-Maxwell source} 
\author{Jos\'e Ocariz\thanks{{\tt ocariz@ciens.ula.ve}}  and H\'ector R
ago\thanks{{\tt rago@ciens.ula.ve}}}
\address{Laboratorio de F\'{\i}sica Te\'orica, Departamento de F\'{\i}sica,
Facultad de Ciencias, Universidad 
de los Andes, M\'erida 5101-A, Venezuela.}
\maketitle
\begin{abstract}
The exterior solution for an arbitrary charged, massive source, is studied
as a static deviation from the Reissner-Nordstr\o m metric. This is reduced to
two coupled ordinary differential equations for the
gravitational and electrostatic potential functions. The homogeneous equations 
are explicitly solved in the particular case $q^2=m^2$, obtaining a multipole
expansion with radial hypergeometric dependence for both potentials. In the 
limiting case of a neutral
source, the equations are shown to coincide with recent results by Bondi
and Rindler.
\end{abstract}
\section{Introduction}
A series of well-known and elegant results in general relativity 
(see references in \cite{BR}), show that the multipole moments of 
an isolated, stationary source 
determine in a complete and unique manner the gravitational 
field on its surrounding 3-space. This has been also extended to the case of 
charged sources, both for static\cite{Hoensenlaers} and stationary \cite{Simon}
cases. Recently however, Bondi and Rindler stated the convenience 
of obtaining precise expressions for the multipole expansion of a fully-
relativistic isolated source by means of a static perturbation defined 
over the Schwarzschild metric. In this way, comparison between newtonian 
gravity and general relativity can be achieved in quantitative terms. In the
same spirit, this work presents an extension {\it \`a la} Bondi and Rindler to 
the Einstein-Maxwell case. Thus, we consider a bounded and static, but 
otherwise arbitrary distribution of charged matter. The assumed static 
character  ensures that the only conserved quantities for such a system are 
the total gravitational mass-energy $m$ and the net charge $q$, as evaluated 
by an observer situated 
well away from the source. It is sensible to expect that such an observer 
would find a space-time not markedly 
different from the Reissner-Nordstr\o m one.  As Bondi and Rindler would 
say, ``the Reissner-Nordstr\o m metric gives bones to the space''. We 
therefore propose a description of the exterior geometry as a small static 
deviation from the Reissner-Nordstr\o m solution. 

This paper is organized as follows: in section {\bf \ref{ecuaciones}}, the 
relevant Einstein-Maxwell equations are presented, and the constraints imposed by the field equations on the perturbation terms are found. In section 
{\bf \ref{analisis}}, the physical implications
of the results are interpreted. The particular cases $q^2=m^2$ and $q=0$
are discussed in detail.

\section{The Einstein-Maxwell field equations}\label{ecuaciones}

In curvature coordinates 
$\left( t,r,\theta ,\varphi \right) $, and taking advantage of the static 
character of the system so as to eliminate the mixed time-space components of 
the metric tensor, and to diagonalize its
pure spatial part by means of a coordinate transformation, we are allowed to 
adopt the following general {\it ansatz} for the line element:
\begin{equation}
\label{metrica}
ds^2=\left( 1+2\alpha \right) pdt^2-\left( 1+2\beta \right)
\frac 1pdr^2-\left( 1+2\gamma \right) r^2d\theta ^2-\left( 1+2\delta \right)
r^2\sin^2\theta d\varphi ^2 ,
\end{equation}
where
\begin{equation}
p\left(r\right)=1-\frac{2m}r+\frac{q^2}{r^2}
\end{equation}
is the usual Reissner-Nordstr\o m metric function, and $\alpha $, $\beta $, 
$\gamma $ and $\delta $ are small functions of the spatial coordinates. On 
the other hand, the vector potential for the electric field will be chosen 
as
\begin{equation}
\label{asuba}
A_a=\left(\frac qr-\Phi \right) \delta ^0_a ,
\end{equation}
with $\Phi $ defined again as a small function of the spatial coordinates. 
It is obvious from (\ref{metrica}) and (\ref{asuba}) that the five-parameter
set of perturbations $\alpha $, $\beta $, $\gamma $, $\delta $ and $\Phi $ 
represent all possible non-spherical terms for the physical fields, and that 
their vanishing brings us back to the spherically symmetric Reissner-
Nordstr\o m solution. This set of functions, defined in a somewhat similar
manner to \cite{BR}, is also completely equivalent to the time-independent
part of Chandrasekhar's
polar perturbations\cite{Chandrasekhar}, when considered as first-order
quantities. Their behaviour  is governed by 
the electrovacuum Einstein-Maxwell field equations,
\begin{eqnarray}
R_{ab}&=&8\pi T_{ab}
\nonumber
\\
&=&2F_{ac}F^{c}_{\ b}+\frac12g_{ab}F_{cd}F^{cd},
\\
\nabla _bF^{ab}&=&0,
\nonumber
\end{eqnarray}
where $F_{ab}=2\nabla _{[a}A_{b]}$ is the electromagnetic Maxwell-Faraday 
tensor.
Since we are interested in the far-field solutions, we can neglect squares and 
products of the perturbations, as well as of their
derivatives, so that the non-vanishing Einstein-Maxwell equations become a set 
of eight partial differential equations, which we transcribe below:

($R_{tt}=8\pi T_{tt}$): 
\begin{eqnarray}
\label{r00}
\frac{\partial ^2\alpha }{\partial r^2}&+&\frac 2r\frac{\partial \alpha }{
\partial r}+\frac{p^{\prime }}{2p}\frac \partial {\partial r}\left( 3\alpha
-\beta +\gamma +\delta \right) +\frac{2q^2}{pr^4}\left( 1+2\alpha \right)
\nonumber
\\
&+&\frac 1{pr^2}\left( \frac{\partial ^2\alpha }{\partial \theta ^2}+\cot
\theta \frac{\partial \alpha }{\partial \theta }+\frac 1{\sin^2\theta }
\frac{\partial ^2\alpha }{\partial \varphi ^2}\right)  \nonumber
\\
&=&-\frac{2q}{pr^2}\frac{\partial \Phi }{\partial r}+\frac{q^2}{pr^4}\left(
1+2\beta \right), 
\end{eqnarray}
($R_{rr}=8\pi T_{rr}$):
\begin{eqnarray}
\label{r11}
\frac{\partial ^2}{\partial r^2}\left( \alpha +\gamma +\delta \right) &-&\frac
2r\frac \partial {\partial r}\left( \beta -\gamma -\delta \right) +\frac{
p^{\prime }}{2p}\frac \partial {\partial r}\left( 3\alpha -\beta +\gamma
+\delta \right) 
\nonumber
\\
&+&\frac 1{pr^2}\left( \frac{\partial ^2\beta }{\partial
\theta ^2}+\cot \theta \frac{\partial \beta }{\partial \theta }+\frac 1
{\sin^2\theta }\frac{\partial ^2\beta }{\partial \varphi ^2}\right)  \nonumber
\\
&=&\frac{2q}{pr^2}\frac{\partial \Phi }{\partial r}-\frac{q^2}{pr^4}\left(
1+2\alpha \right), 
\end{eqnarray}
($R_{\theta \theta }=8\pi T_{\theta \theta }$):
\begin{eqnarray}
\label{r22}
\frac{\partial ^2\gamma }{\partial r^2}&+&\frac{p^{\prime }}p\frac{\partial
\gamma }{\partial r}+\frac 1r\frac \partial {\partial r}\left( \alpha -\beta
+3\gamma +\delta \right) -\frac 2{pr^2}\left( \beta -\gamma \right) 
\nonumber
\\
&+&\frac 1 {pr^2}\left[ \frac{\partial ^2}{\partial \theta ^2}\left( \alpha +
\beta
+\delta \right) +\cot \theta \frac \partial {\partial \theta }\left( 2\delta
-\gamma \right) +\frac 1{\sin^2\theta }\frac{\partial ^2\gamma }{\partial
\varphi ^2}\right]  \nonumber
\\
&=&\frac{2q}{pr^2}\left( \frac{\partial \Phi }{\partial r}+\frac q{r^2}\alpha
\right), 
\end{eqnarray}
($R_{\varphi \varphi }=8\pi T_{\varphi \varphi }$):
\begin{eqnarray}
\label{r33}
\frac{\partial ^2\delta }{\partial r^2}&+&\frac{p^{\prime }}p\frac{\partial
\delta }{\partial r}+\frac 1r\frac \partial {\partial r}\left( \alpha -\beta
+\gamma +3\delta \right) -\frac {2\left( \beta -\gamma \right) }{pr^2}
\nonumber
\\
&+&\frac
1{pr^2}\left[ \frac{\partial ^2\delta }{\partial \theta ^2}+\cot \theta
\frac \partial {\partial \theta }\left( \alpha +\beta -\gamma +2\delta
\right) +\frac 1{\sin^2\theta }\frac{\partial ^2}{\partial \varphi ^2}
\left( \alpha +\beta +\gamma \right) \right]  \nonumber
\\
&=&\frac{2q}{pr^2}\left( \frac{\partial \Phi }{\partial r}+\frac q{r^2}\alpha
\right), 
\end{eqnarray}
($R_{r\theta }=8\pi T_{r\theta }$):
\begin{eqnarray}
\label{r12}
\frac{\partial ^2}{\partial r\partial \theta }\left( \alpha +\delta \right)
&+&\cot \theta \frac \partial {\partial r}\left( \delta -\gamma \right) +
\frac{p^{\prime }}{2p}\frac \partial {\partial \theta }\left( \alpha -\beta
\right) -\frac 1r\frac \partial {\partial \theta }\left( \alpha +\beta
\right) =-\frac{2q}{pr^2}\frac{\partial \Phi }{\partial \theta },
\end{eqnarray}
($R_{r\phi }=8\pi T_{r\phi }$):
\begin{eqnarray}
\label{r13}
\frac{\partial ^2}{\partial r\partial \varphi }\left( \alpha 
+\gamma \right) +
\frac{p^{\prime }}{2p}\frac \partial {\partial \varphi }\left( \alpha -\beta
\right) -\frac 1r\frac \partial {\partial \varphi }\left( \alpha +\beta
\right) =-\frac{2q}{pr^2}\frac{\partial \Phi }{\partial \varphi },
\end{eqnarray}
($R_{\theta \varphi }=8\pi T_{\theta \varphi }$):
\begin{equation}
\label{r23}
\frac{\partial ^2}{\partial \theta \partial \varphi }\left(
\alpha +\beta \right) -\cot \theta \frac \partial {\partial \varphi }\left(
\alpha +\beta \right) =0,
\end{equation}
($\nabla_aF^{ta}=0$):
\begin{eqnarray}
\label{f0a}
\frac{\partial ^2\Phi }{\partial r^2}&+&\frac 2r\frac{\partial \Phi }{\partial
r}+\frac 1{pr^2}\left( \frac{\partial ^2\Phi }{\partial \theta ^2}+\cot
\theta \frac{\partial \Phi }{\partial \theta }+\frac 1{\sin^2\theta }
\frac{\partial ^2\Phi }{\partial \varphi ^2}\right) +\frac q{r^2}\frac
\partial {\partial r}\left( \alpha +\beta -\gamma -\delta \right) =0.
\end{eqnarray}

\subsection*{Resolution}
Since all static and axisymmetric solutions to the Einstein-Maxwell 
equations belong to Weyl's {\it electrovac} class\cite{Kramer}, it is convenient 
to define the five perturbation functions in such a way as to avoid the 
repetition of these known results. On the other hand, it is also convenient 
to exclude the possibility that those perturbations arise simply as a 
consequence of an inappropriate choice of coordinate origin, in which case 
even the spherically symmetric solution would possess such non-zero metric 
functions. Both exclusions may be easily achieved in the selected coordinate 
system, by means of an expansion in spherical harmonics in the usual way:
\begin{equation}
\label{expansion}
\alpha \left( r,\theta ,\varphi \right) =\stackrel{\infty }
{\stackunder{n=2}{\sum }}\ \stackrel{n}{\stackunder{\stackunder{(l\neq 0)}
{l=-n}}{\sum }}\alpha _{nl}\left( r\right) Y_n^l\left( \theta ,\varphi 
\right) ,
\end{equation}
and other similar expansions for the remaining functions $\beta $, 
$\gamma $, $\delta $ and $\Phi $. It is important to notice that the chosen set 
of indexes $n,l$ explicitly imposes the above mentioned restrictions, for $n=0$
represents the spherically symmetric terms, which are already fully present 
in the background Reissner-Nordstr\o m curvature; the $l=0$ set corresponds 
to the axisymmetric multipoles; and finally, a
non-vanishing $n=1$, $l=\pm 1$ term simply means a shifted choice of 
coordinate origin.

Keeping this in mind, we notice that the field equations (\ref{r23}), 
(\ref{r13}) and (\ref{r12}) impose the following relations between the
perturbation functions
\begin{eqnarray}
\beta &=&-\alpha ,\nonumber
\\
\frac{\partial \gamma }{\partial r}&=&-\left[ \frac{2q}{pr^2}\Phi
+\frac{\partial \alpha }{\partial r}+\frac{p^{\prime }}p\alpha \right] ,
\nonumber
\\
\label{vinculos}
\delta &=&\gamma ;
\end{eqnarray}
whose inclusion in the remaining field equations show that (\ref{r00}) and
(\ref{r11}), as well as (\ref{r22}) and (\ref{r33}) become identical, 
leaving us with just three independent equations. Using the expansions 
(\ref{expansion}), and adopting the notation 
$\tilde{\alpha }\left( r\right) $, $\tilde{\gamma }\left( r\right) $ and 
$\tilde{\Phi }\left( r\right) $ to represent the set of radial coefficients  
$\alpha _{nl}$, $\gamma _{nl}$ and $\Phi _{nl}$, all of them with the same 
index $n$, these equations reduce to a system of two ordinary differential 
equations on $\tilde{\alpha }$ and $\tilde{\Phi }$, namely:
\begin{mathletters}
\label{ord}
\begin{eqnarray}
\label{alfaord}
\frac{d^2\tilde{\alpha }}{dr^2}+\left[ \frac{p^{\prime }}
p+\frac 2r\right] \frac{d\tilde{\alpha }}{dr}+\left[ \frac{2q^2}{pr^4}
- \frac{p^{\prime 2}}{p^2}-\frac{n\left( n+1\right) }{pr^2}\right] 
\tilde{\alpha }&=&\frac{2q}{pr^2}\left[ \frac{p^{\prime }}p
\tilde{\Phi }-\frac{d\tilde{\Phi }}{dr}\right] ,
\\
\label{fiord}
\frac{d^2\tilde{\Phi }}{dr^2}+\frac 2r\frac{d\tilde{\Phi } 
}{dr}+\left[ \frac{4q^2-n\left( n+1\right) r^2}{pr^4}\right] 
\tilde{\Phi }&=&-\frac{2q}{r^2}\left[ \frac{d\tilde{\alpha }}{dr}+
\frac{p^{\prime }}p \tilde{\alpha }\right] ,
\end{eqnarray}
\end{mathletters}
plus an algebraic relation for $\tilde{\gamma }$
\begin{equation}
\label{gamaord}
\frac{d^2\tilde{\alpha }}{dr^2}+2\left[ \frac{p^{\prime }}p+\frac 1r
\right] \frac{d\tilde{\alpha }}{dr}+\frac 2{r^2}\left[ \frac{q^2}
{pr^2}-1\right] \tilde{\alpha }+\frac{\left( n+2\right) \left(
n-1\right) }{pr^2}\tilde{\gamma }=-\frac{4q}{pr^3}\left[ \tilde{\Phi }
+r\frac{d\tilde{\Phi }}{dr}\right] .
\end{equation}
In this way, a simultaneous solution $\left[ \tilde{\alpha }\left( r
\right) ;\tilde{\Phi }\left( r\right) \right] $ of (\ref{ord}), would 
determine, via (\ref{gamaord}), 
(\ref{vinculos}) and the expansions (\ref{expansion}), the Einstein-Maxwell
fields (\ref{metrica}) and (\ref{asuba}).

The so obtained field equations show the convenience of the
assumed linearization in a Reissner-Nordstr\o m background, since 
the absence of cross-terms  has eliminated the $l$-dependence in (\ref{ord}) 
for each ($2n$)-set of multipoles with the same
index $n$, thus recovering the well-known fact, that it is always
possible to restrict ourselves to an axisimmetric dependence,
completely determined by the index $n$.

In the limiting case of null electric field, $q=\Phi =0$, our set of 
independent equations (\ref{ord}) reduces only to one 
differential equation on the gravitational perturbation 
$\tilde{\alpha }\left( r\right) $, namely
\begin{equation}
\label{alfa0}
\frac{d^2\widetilde{\alpha }}{dr^2}+\frac{2\left( r-m\right) }{%
r\left( r-2m\right) }\frac{d\widetilde{\alpha }}{dr}-\frac{4m^2+n\left(
n+1\right) \left( r-2m\right) r}{r^2\left( r-2m\right) ^2}
\widetilde{\alpha }=0.
\end{equation} 
The simple transformation from our radial curvature coordinate $r$ to
the radial isotropic coordinate $\rho $ through  a new
dimensionless radial parameter $x$,
\begin{equation}
x=\left( \dfrac{m}{2\rho}\right) ^2\ \rightarrow\ r=\frac{m}{2}\
frac{\left(1+\sqrt{x}\right)^2}{\sqrt{x}},
\end{equation}
converts (\ref{alfaord}) into
\begin{equation}
\left(1-x\right)^2x^2\frac{d^2\tilde{\alpha}}{dx^2}+\frac{1}{2}\left(1-3x\right)
\left(1-x\right)x\frac{d\tilde{\alpha}}{dx}-\left[4x+\frac{1}{4}n\left(n+1\right)
\left(1-x\right)^2\right]\tilde{\alpha}=0,
\end{equation}
for $\left\{ 0\leq x<1\right\} $ corresponding
to the region $\left\{ 2m<r\leq \infty \right\} $, which is identical to the 
one obtained in \cite{BR}. Obviously, the
form of (\ref{alfa0}), which is an homogeneous differential equation with 
three regular singularities, ensures that it is possible to obtain exact 
solutions directly from it, but their demonstrated equivalence with the 
Bondi-Rindler
solutions doesn't make it worth while.

\section{Analysis of the Equations}\label{analisis}

We would like to emphasize that the electric and gravitational fields are
completely determined by (\ref{ord}), which are a coupled system of linear 
differential equations. The general 
relativistic coupling between these fields is obvious, since both equations 
possess on their right hand side an inhomogeneous term. Therefore, the 
general solution of (\ref{ord}) can be split into an 
homogeneous part for each function $\tilde{\alpha }$ and $\tilde{\Phi }$, 
plus the inhomogeneous, coupled terms. 

Furthermore, (\ref{alfaord}) and (\ref{fiord}) are elliptic
equations, and hence they define a well-posed boundary problem, whose
boundary conditions must be imposed as physically meaningful constraints.
In addition, their elliptic character offers an extra
advantage, for the required boundary conditions will have to be satisfied
not only by the entire solutions $\left[ \tilde{\alpha }\left( r
\right) ;\tilde{\Phi }\left( r\right) \right] $, but also by their
homogeneous parts $\left[ \tilde{\alpha }_0\left( r
\right) ;\tilde{\Phi }_0\left( r\right) \right] $, and this can be simply
checked out by means of the simpler set of homogeneous equations, given
by the left hand sides of (\ref{ord}).

The most obvious boundary condition was already assumed in the definition of 
the perturbations, for these must vanish at large distances from the 
distribution in order to recover the Reissner-Nordstr\o m solution. In this 
way, it is straightforward to verify that both $\tilde{\alpha }$ 
and $\tilde{\Phi }$ 
are determined by two arbitrary integration constants, and therefore,
for both fields, each 
set of harmonics possesses the same multiplicity of solutions 
as the classical study of independent gravitational and electrostatic 
potentials. Likewise, in each case only one of the two possible 
solutions will vanish as they approach infinity. On the other hand, the second 
boundary condition should be imposed by the junction with a convenient 
interior solution; but its arbitrary character would make it very difficult to 
define in a tractable way. Fortunately, the metric (\ref{metrica}) 
inherits a physical feature from the chosen background curvature, at 
least for the general cases $q^2\leq m^2$, since the analytic extension of 
the Reissner-Nordstr\o m solution 
describes a black hole, with an
event horizon situated at $r_+=m+\sqrt{m^2-q^2}$, and its 
``no-hair'' property brings a suitable boundary condition. This can 
be easily shown analytically,
and without invalidating the generality of the above remark, if we restrict 
ourselves to the simpler case $q^2=m^2$. The homogeneous parts 
of (\ref{ord}) can be written, in terms of a convenient
dimensionless radial parameter $x=\dfrac{m}{r}$, as follows:
\begin{mathletters}
\label{x}
\begin{eqnarray}
\label{alfax}
x^2\left( 1-x\right) ^2\frac{d^2\tilde{\alpha }_0}{dx^2}+2x^2\left( x-1
\right) \frac{d\tilde{\alpha }_0}{dx}-\left[ 2x^2+n\left(
n+1\right) \right] \tilde{\alpha }_0&=&0 ,
\\
\label{fix}
x^2\left( 1-x\right) ^2\frac{d^2\tilde{\Phi }_0}{dx^2}+\left[
4x^2-n\left( n+1\right) \right] \tilde{\Phi }_0&=&0  .
\end{eqnarray}
\end{mathletters}
In this way, the $r\rightarrow \infty $ boundary condition is now located at
the finite value $x=0$, and the event horizon for the maximal charged black 
hole at $x=1$, so that the outside space is covered by the
region $\left\{0<x\leq 1\right\}$. As both equations now possess three
regular singularities at $x=0,1,\infty $, the former are granted 
to have solutions
with an hypergeometric dependence given by some functions $h\left( x\right) $, 
defined as follows:
\begin{equation}
\label{alfafi}
\begin{array}{ccc}
\tilde{\alpha }_0=x^{p_1}\left( 1-x\right) ^{q_1}h_1\left( x\right)
&,&
\tilde{\Phi }_0=x^{p_2}\left( 1-x\right) ^{q_2}h_2\left( x\right);
\end{array}
\end{equation}
whose insertion in (\ref{x}) brings up the following pairs
of solutions for the exponents $p,q$, suitably expressed in terms of 
an index $j=n+\frac{1}{2}$:
\begin{equation}
\label{pq}
p_1,p_2=\left\{ \begin{array}{c}n+1\\n\end{array}\right.
\ ,\ 
q_1=-\frac{1}{2}\pm \sqrt{j^2+2}
\ ,\ 
q_2=+\frac{1}{2}\pm \sqrt{j^2-4}.
\end{equation}
It is then obvious that only the solutions with $p$ equal to $n+1$ will
satisfy the asymptotic vanishing condition. On the other hand, the $q$'s
with positive root must be discarded, for they would lead to vanishing
perturbations at the event horizon $x=1$, showing a spherically symmetric 
configuration at the finite radial value $r=m$; this situation would be 
in flagrant contradiction with Birkhoff's theorem. In this way, we are led
 to a unique set of choices; with these, (\ref{x}) can be cast 
into the canonical form of standard hypergeometric 
equations
\begin{mathletters}
\label{maxdef}
\begin{eqnarray}
x\left( 1-x\right) \frac{d^2h_1}{dx^2}
+\left[ 2\left( n+1\right) -x\left( 2n+3+2\sqrt{j^2+2}\right) \right]
\frac{dh_1}{dx} \nonumber
\\
\label{alfamaxdef}
+\left[ 2\left( n+1\right) \sqrt{j^2+2}-2j
\left( n+1\right) \right] h_1=0 ,
\\
x\left( 1-x\right) \frac{d^2h_2}{dx^2}
+\left[ 2\left( n+1\right) -x\left( 2n+3+2\sqrt{j^2-4}\right) \right]
\frac{dh_2}{dx} \nonumber
\\
\label{fimaxdef}
+\left[ 2\left( n+1\right) \sqrt{j^2-4}-2j\left( n+1\right) \right] h_2=0 ;
\end{eqnarray}
\end{mathletters}
whose integrals are given by the hypergeometric functions
\begin{equation}
\label{h}
h_1\left( x\right) =H\left( a_1,b_1,2n+2;x\right)
\ , \
h_2\left( x\right) =H\left( a_2,b_2,2n+2;x\right);
\end{equation}
with the $a$'s and $b$'s  defined according to
\begin{equation}
\label{ab}
\left\{
\begin{array}{ccc}
a_1+b_1&=&2\left( n+1-\sqrt{j^2+2}\right)
\\
a_1b_1&=&2\left( n+1\right) \left(j-\sqrt{j^2+2}\right)
\end{array}
\right.
\ ,\
\left\{
\begin{array}{ccc}
a_2+b_2&=&2\left( n+1-\sqrt{j^2-4}\right)
\\
a_2b_2&=&2\left( n+1\right) \left(j-\sqrt{j^2-4}\right) .
\end{array}
\right. 
\end{equation}
Although in both cases $a$ and $b$ are complex numbers, the regularity of $h_1$
and $h_2$ is completely ensured by their hypergeometric definition
\begin{equation}
\label{hyper}
H\left( a,b,c;x\right)=1+\frac{ab}{c}\frac{x}{1!}+
\frac{a(a+1)b(b+1)}{c(c+1)}\frac{x^2}{2!}+\cdots 
=\stackrel{\infty }{\stackunder{n=0}{\sum }}\frac{(a)_n(b)_n}{(c)_n}
\frac{x^n}{n!}
\end{equation}
since all $a$ and $b$ solutions of (\ref{ab}) are 
complex conjugates,  the functions in (\ref{hyper})
are always real-valued. It is then obvious that $h_1$ and 
$h_2$ possess a regular behaviour in the $\left[ 0;1\right] $ range; moreover, 
their hypergeometric dependence ensures also that they are monotonously 
increasing functions in that same range, as can be easily seen by keeping 
in mind their vanishing at $x=0$, and evaluating them and their first 
derivatives at $x=1$. So, the radial dependence of $\tilde{\alpha }$ and 
$\tilde{\Phi }$ will be mostly dominated by the $x^p\left( 1-x\right) ^q$ 
terms in (\ref{alfafi}): For $x\rightarrow 0$, (\ref{pq}) shows that the 
asymptotic behaviour will be determined by the $x^{n+1}$ term, thus inducing 
a radial dependence given by
\begin{equation}
\label{asintotico}
\widetilde{\alpha },\widetilde{\Phi }\sim \frac{1}{r^{n+1}},
\end{equation}
which corresponds to the classical coulombian decreasing rate; this also 
shows that the coupling terms between both fields, i.e. those who are 
non-coulombian contributions, asymptotically vanish with greater powers of 
the radial coordinate, and only the harmonic (\ref{asintotico}) part of the
fields will be perceptible. On the other side, both $\tilde{\alpha }$ and 
$\tilde{\Phi }$ diverge
in the $x\rightarrow 1$ limit, since their radial dependence will be 
dominated by the $\left( 1-x\right) ^q$ terms, and the $q$'s defined in
(\ref{pq}) are all negative for the permitted values of $n$. This divergent
boundary condition at $x=1$ is in agreement with the ``no-hair'' theorem
for charged black holes, since it ensures that there can be no bounded 
static perturbations of the analytically extended electrovac solutions on 
the event horizon. It also shows that in the $x\sim 1$ strong-field region,
the coupling between both fields dominates over their coulombian
parts, given by (\ref{asintotico}).
The homogeneous functions  $\left[ \tilde{\alpha}_0;\tilde{\Phi}_0\right] $ 
so obtained also offer an algorithmic way to compute the complete inhomogeneous 
field equations, since these functions
can now be substituted into the right hand sides of (\ref{ord})
in order to find some first inhomogeneous functions 
$\left[ \tilde{\alpha}_i;\tilde{\Phi}_i\right]$.
Iterating this procedure leads to more accurate inhomogeneous functions,
which, when added to the homogeneous ones, will bring exact solutions 
to the field equations. Anyway, it must be born in mind that these
functions might not be exact solutions of the field equations out of 
the weak-field condition, where their smallness condition has
to be relaxed.
Nevertheless, as discussed in \cite{BR}, the ellipticity of
  (\ref{ord}) ensures that no solutions other than these can exist
outside of a sphere which contains all the sources.   

\acknowledgements
It is a pleasure to express our gratitude to L.A. Nu\~nez for  
enlightening discussions. We would also like to thank our referees for
their valuable comments.
J.O. is indebted with the {\it Fundaci\'on para el
Desarrollo de la Ciencia y la Tecnolog\'{\i}a FUNDACITE M\'erida}, 
for their kind support throughout his undergraduate thesis. This 
work was partially financed by the {\it Consejo de Desarrollo 
Cient\'{\i}fico, Human\'{\i}stico y Tecnol\'ogico C.D.C.H.T.}, grant number 
$C-606-93-CC(05)-F$.


\begin{thebibliography}{99}
\bibitem{BR}Bondi H., Rindler W., {\it Gen. Rel. Grav. 23}, 487 (1991).
\bibitem{Hoensenlaers}Hoensenlaers C., {\it Prog. Theor. Phys. 55}, 466 (1976).
\bibitem{Simon} Simon W., {\it J. Math. Phys. 25}, 1035 (1984).
\bibitem{Chandrasekhar}Chandrasekhar S., {\it The Mathematical Theory of
Black Holes}, (Clarendon, 1983).
\bibitem{Kramer}Kramer D., Stephani H., Herlt E., MacCallum M., {\it Exact 
Solutions of Einstein's 
Field Equations}, (Cambridge University Press, 1980).
\end{thebibliography}
\end{document}